\documentclass[11pt]{article}

\usepackage{graphicx}
\usepackage[abs]{overpic}
\usepackage{subfigure}
\usepackage{url}
\usepackage{multirow}

\usepackage[margin=1in]{geometry}
\usepackage[inline]{enumitem}

\newenvironment{inlineEnum}[1][and]
{\begin{enumerate*}[label=(\alph*),before=\unskip{: }, itemjoin={{; }}, itemjoin*={{, #1 }}]{}}
{\end{enumerate*}}

\newcommand{\bb}{$b\bar{b}$}
\newcommand{\ppbar}{$p\bar{p}$}
\newcommand{\Hbb}{$H \rightarrow b\bar{b}$}
\newcommand{\bbbar}{$b\bar{b}$}
\newcommand{\ggH}{$gg \rightarrow H$}
\newcommand{\invfb}{\mbox{fb$^{-1}$}}

\newcommand{\WHlvbb}{\mbox{$WH \rightarrow \ell \nu b\bar{b}$}}
\newcommand{\ZHllbb}{\mbox{$ZH \rightarrow \ell \ell  b \bar{b}$}}
\newcommand{\ZHvvbb}{\mbox{$ZH \rightarrow \nu \nu  b \bar{b}$}}
\newcommand{\mH}{$m_{H}$}

\newcommand{\gevcc}{GeV/$c^2$}

\newcommand{\HWW}{$H \rightarrow WW$}

\newcommand{\ttbar}{\mbox{$t\bar{t}$}}

\newcommand{\Et}{\mbox{$E_T$}}

\newcommand{\met}{\mbox{$\protect \raisebox{0.3ex}{$\not$}\Et$}}

\newcommand{\jetprob}{{\sc jetprob}}
\newcommand{\secvtx}{{\sc secvtx}}

\title{Search for the Standard Model Higgs boson in final states with $b$ quarks at the Tevatron}

\author{Karolos Potamianos\thanks{On behalf of the CDF and D\O\ collaborations}\\
       Purdue University\\
       Fermilab, Batavia, IL, USA\\
       E-mail: \url{karolos.potamianos@cern.ch}}
       \date{
  	2011 Europhysics Conference on High Energy Physics-HEP 2011\\
	 July 21-27, 2011\\ Grenoble, Rh\^one-Alpes France \\\vspace*{2em} \today
       }

\begin{document}
\maketitle

\begin{abstract}
We present the result of searches for a low mass Standard Model Higgs boson produced in association with a $W$ or a $Z$ boson at a center-of-mass energy of $\sqrt{s}=$1.96~TeV with the CDF and D0 detectors at the Fermilab Tevatron collider. The search is performed in events containing one or two $b$ tagged jets in association with either two leptons, or one lepton and an imbalance in transverse energy, or simply a large imbalance in transverse energy. Datasets corresponding to up to 8.5~fb$^{-1}$ of integrated luminosity are considered in the analyses. These are the most powerful channels in the search for a low mass Higgs boson at the Tevatron. Recent sensitivity improvements are discussed. For a Higgs mass of $115$~\gevcc, the expected sensitivity for the most sensitive individual analyses reaches 2.3 times the SM prediction at 95\% confidence level (C.L.), with all limits below 5 times the SM. Additionally, a $WZ/ZZ$ cross-section measurement is performed to validate the analysis techniques deployed for searching for the Higgs. 
\end{abstract}

\section{Introduction}
\vspace*{-.1in}

\begin{figure}[t]
\centering
\begin{tabular}{lr}
 \begin{overpic}[width=.4\linewidth]{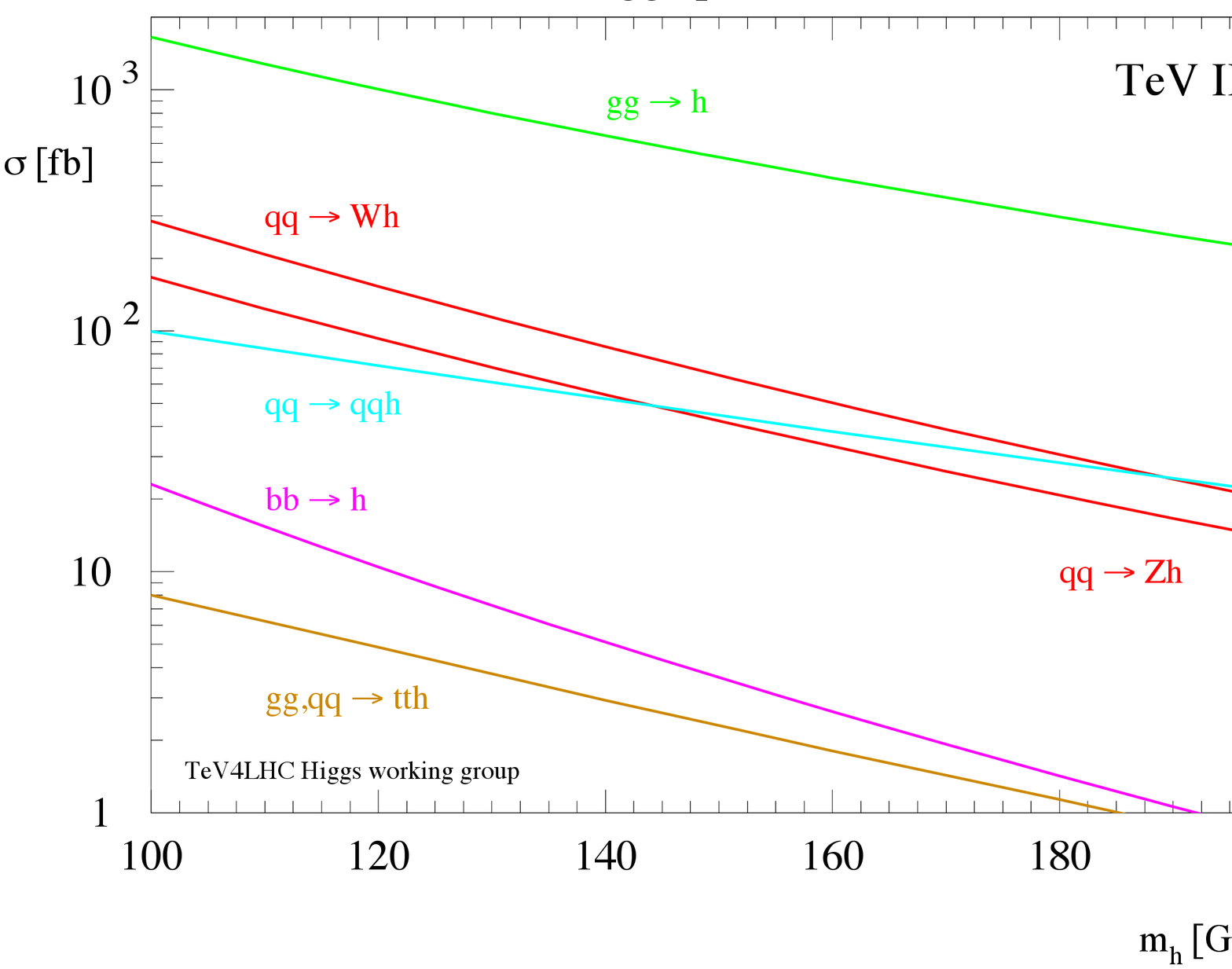}\put(170,110){(a)}\end{overpic}
 & 
\hspace*{.5in} \begin{overpic}[bb=1 -9 453 354,width=.4\linewidth]{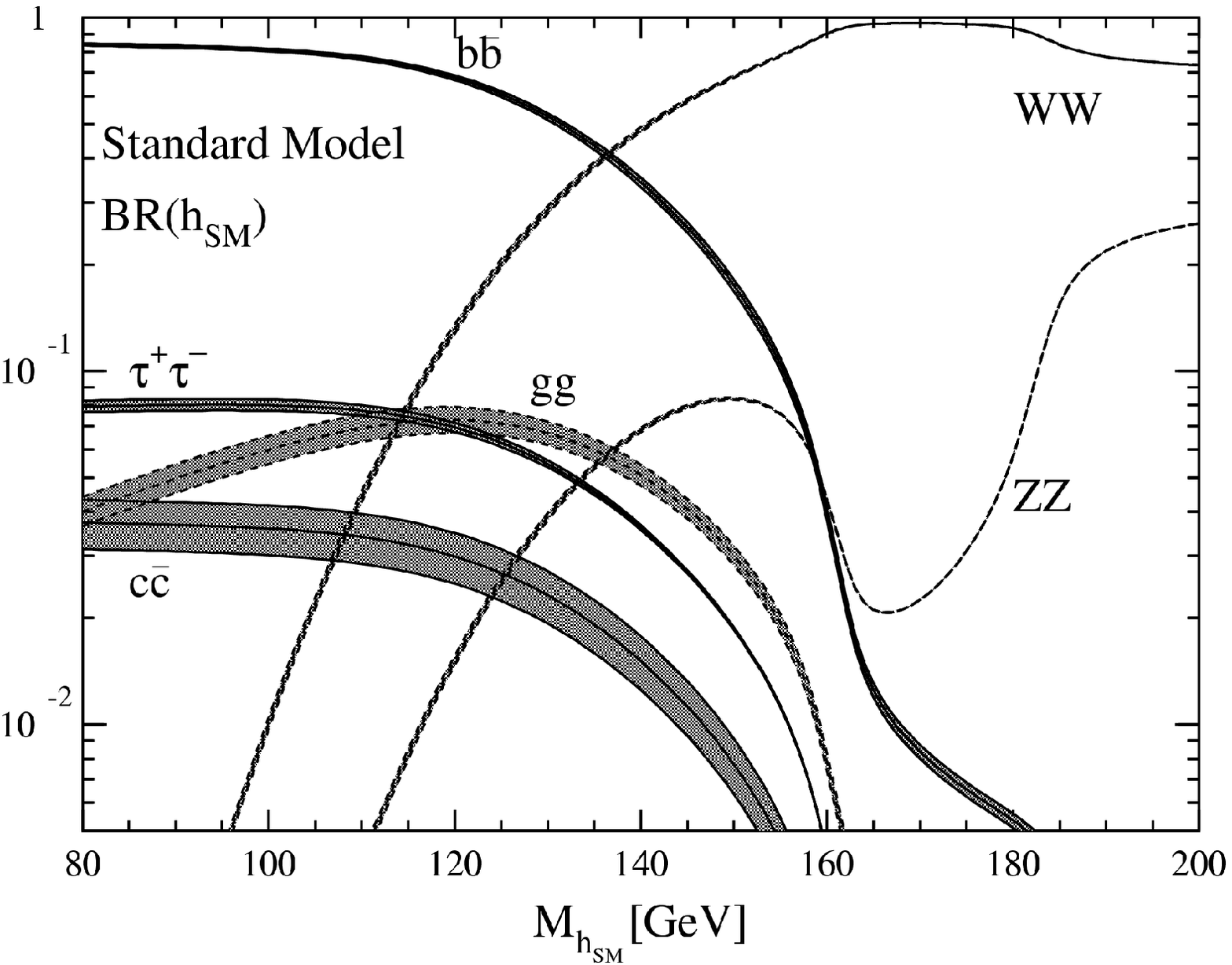}\put(150,110){(b)}\end{overpic}
\end{tabular}
\caption{Higgs production cross-sections (a) and branching ratios (b) at the Tevatron (\ppbar,$\sqrt{s} = 1.96$~TeV).\vspace*{-.05in}}
\label{fig:SMHiggsXSecAndDecay}
\end{figure}

Understanding electroweak symmetry breaking has been a major goal of the high energy physics community for several decades. 
The Higgs mechanism\footnote{In full: Brout-Englert-Higgs-Hagan-Guralnik-Kibble (BEHHGK) mechanism.}~\cite{ref:BEHHGK} proposed in 1964 added the Higgs boson to the standard model (SM) of particle physics; it has yet to be observed. 
Direct searches at LEP have placed a lower limit of $114.4$~\gevcc\ on the SM Higgs boson mass (\mH) at 95\% confidence level (C.L.), while precision electroweak measurements place an indirect limit of $ m_H < 158$~\gevcc\ at 95\% C.L.~\cite{ref:LEPEWWG}.
Figure~\ref{fig:SMHiggsXSecAndDecay} shows the production and decay modes of the SM Higgs boson ($H$) at the Tevatron \ppbar\ collider ($\sqrt{s} = 1.96$~TeV).
The recent observation of single top~\cite{ref:single_top} and diboson~\cite{ref:diboson} production in semi-leptonic decays have prepared the way for Tevatron experiments to probe processes with  sub-picobarn cross sections, among which is the SM Higgs boson, now a central part of the Tevatron program.
We present the status of direct searches for a low mass SM Higgs boson using up to $8.5$~\invfb\ of data collected by the CDF~II and D\O\ detectors~\cite{ref:CDFD0}.

\vspace*{-.1in}
\section{Low mass Higgs searches}
\vspace*{-.1in}

The search for the SM Higgs boson is challenging due to the small signal expectation and the large backgrounds\footnote{At the Tevatron, dijet QCD events are produced at a rate ten orders of magnitude higher than the SM Higgs boson.}. The search strategy at the Tevatron is based on a {\it divide and conquer} approach: since no single signature has sufficient sensitivity, we divide the search into many different channels. Each channel exploits a decay mode using dedicated triggers and analysis techniques.

According to Figure~\ref{fig:SMHiggsXSecAndDecay}, 
\Hbb\ is the dominant mode for $m_H < 135$~\gevcc\ (low mass) while \HWW\ dominates at high mass. Because of the overwhelming irreducible QCD background, it is necessary to seek a striking signature to identify the SM Higgs boson. At low mass, we focus on processes where it is produced in association with a $W$ or $Z$ boson, $VH$, while at high mass, the decay products of the $W$ allow probing of the five times more likely direct production \ggH.

The main low mass channels are dedicated to events with zero (\met+\bb)~\cite{ref:cdf_vvbb,ref:dzero_vvbb}, one ($\ell\nu$\bb)~\cite{ref:cdf_lvbb,ref:dzero_lvbb} or two ($\ell\ell$\bb)~\cite{ref:cdf_llbb,ref:dzero_llbb} identified leptons from the \ZHvvbb, \WHlvbb, and \ZHllbb\ decay modes. The background consists of QCD multi-jet production faking the above signatures, the production of a $W$ or $Z$ in association with jets, single top and top pair productions, and diboson production. 
These results are combined with the high mass and additional low mass channels to determine the sensitivity of the Tevatron experiments to the SM Higgs boson~\cite{ref:tevHiggsComb2011}.

\section{Analysis techniques}
\vspace*{-.1in}

After a cut-based event selection, similar for both experiments and depending on the search channel, most analyses resort to multivariate analysis to reduce the multi-jet background; $b$ quark identification further improves the signal-to-background ratio. Finally, several discriminants -- either neural networks (NN)~\cite{ref:cdf_vvbb,ref:cdf_lvbb}, boosted decision trees (BDT)~\cite{ref:dzero_vvbb,ref:dzero_lvbb}, or ensembles (forests) thereof~\cite{ref:cdf_llbb,ref:dzero_llbb} -- are trained to maximize the sensitivity. 

Because the individual analyses use multivariate techniques to exploit the information in each event, it is crucial to check that all the inputs to these techniques, as well as their outputs, are well described by the background models. We perform these checks comparing the background model to the data in various control regions. Each of these regions is defined to check the modeling of a major background component. The CDF and D\O\ collaborations have built confidence in detector modeling, and multivariate techniques have been successfully used in recent observations~\cite{ref:single_top,ref:diboson}.


\subsection{$b$ quark identification}
Because \Hbb\ is the dominant decay mode for a low mass Higgs boson, it is crucial to identify the jets originating from these two $b$ quarks. 

The CDF analyses~\cite{ref:cdf_vvbb,ref:cdf_lvbb,ref:cdf_llbb} require that at least one jet in the event be identified as originating from a $b$ quark (tagged) by the \secvtx~\cite{ref:cdf_secvtx} algorithm, which identifies $b$ quarks by fitting tracks displaced from the primary vertex.
The second jet in the event can be either
\begin{inlineEnum}[or]
\item not identified as originating from a $b$ quark
\item tagged by the \secvtx\ algorithm
\item tagged by the \jetprob~\cite{ref:cdf_jetprob} algorithm, which uses the impact parameter of the tracks to determine the  probability that all tracks contained in a jet originate from the primary vertex, and identifies $b$ quarks by requiring a low value for that probability
\item tagged by a neural-network-based tagging algorithm~\cite{ref:cdf_lvbb} that identifies $b$ quarks from combined information on displaced vertices, displaced tracks, and low $p_T$ muons
\end{inlineEnum}.

The D\O\ analyses~\cite{ref:dzero_vvbb,ref:dzero_lvbb,ref:dzero_llbb} deploy a BDT algorithm~\cite{ref:dzero_btagging} designed to discriminate $b$ from light ($u$, $d$, $s$, $g$) jets to select events with one or more $b$ quark candidates. The algorithm includes information relating to the lifetime of the hadrons in the jet and results in a discrimination between $b$ and light jets. 
The analysis samples are divided into two channels, where exactly one (single tag) or two (double tag) of the leading jets are above a certain value of the BDT output, $L_b$. In two analyses~\cite{ref:dzero_vvbb,ref:dzero_lvbb}, $L_b$ is also an input to the final discriminant. 

\subsection{Combination of multiple triggers}

Physics experiments at hardon colliders heavily rely on a trigger system to select interesting collision events. Dedicated trigger paths meet specific physics goals. Combining multiple paths maximizes acceptance by collecting events that did not fire the dedicated trigger of a given analysis, but which are nonetheless worth investigating. This combination can be performed either 
by an \emph{a priori} partitioning of the events into orthogonal trigger samples, and checking whether the trigger assigned to the sample fired or not (CDF  $WH\to\ell\nu$\bbbar~\cite{ref:cdf_lvbb}), or
by defining a new path, consisting of a logical OR of the trigger paths~\cite{ref:cdf_vvbb,ref:dzero_lvbb}.
The trigger efficiency must, however, be adequately parameterized to accurately account for the effect of the trigger in the Monte Carlo simulated events. By design, partitioning the events into orthogonal trigger samples requires no additional treatment with respect to the one-trigger approach. 
	In the case of the logical combination, the CDF $VH\to\met$\bbbar\ analysis~\cite{ref:cdf_vvbb} models the trigger efficiency with a neural-network based parameterization, which combines 9 (14) input variables for events with two (three) jets;
the D\O\ $WH\to\ell\nu$\bbbar~\cite{ref:dzero_lvbb} uses a parameterization based on the lepton $\eta$ and $\phi$, and the jet $p_T$.

\subsection{Increased acceptance}

The improvement in modeling the trigger efficiency allows to include more triggers. In the CDF $VH\to\met$\bbbar\  analysis, it allows to significantly relax the kinematic cuts, gaining 40\% in signal acceptance (and 7-10\% in the expected limit). 

To increase acceptance to leptons, the CDF $ZH\to\ell\ell$\bbbar\ analysis~\cite{ref:cdf_llbb} deploys a multivariate ($NN$) lepton identification which uses the lepton $p_T$, $\eta$, $\phi$, $E_{EM}$, $E_{HAD}$, as well as $\Delta R(\ell,j)$, the track $\chi^2$, impact parameter, isolation, and the number of hits in the silicon detector, which provides a 20\% improvement over to a cut-based selection; the CDF $WH\to\ell\nu$\bbbar\ analysis includes a new loose electron and isolated track category, included separately, increasing the sensitivity by $5\%$.


\subsection{Improved discrimination}

The CDF $ZH\to\ell\ell$\bbbar\ analysis~\cite{ref:cdf_llbb} plots the discriminant in three regions defined using two $NN$s to separate the events according to their $\ttbar$, light and heavy flavor score. Constraining the backgrounds in the first two regions improves the sensitivity by $8\%$ over the case with no separation.

The D\O\ $ZH\to\nu\nu$\bbbar\ and $WH\to\ell\nu$\bbbar\ analyses~\cite{ref:dzero_vvbb,ref:dzero_lvbb} improve the sensitivity of their discriminant by 5-10\% by using the output for the BDT tagger as an input to their discriminants.

%

\vspace*{-.1in}
\section{Reinterpretation as a search for $WZ/ZZ$}
\vspace*{-.1in}

The D\O\ $VH\to\met$\bbbar\ analysis~\cite{ref:dzero_vvbb} reinterprets its data as a search for $WZ$ and $ZZ$ production; it uses a BDT trained using the $WZ+ZZ$ signal to measure \mbox{$\sigma_{WZ+ZZ} = 6.9 \pm 1.3 {\rm(stat)} \pm 1.8 {\rm(syst)}$~pb} (consistent with the predicted SM value of $4.6$ pb) with an expected (observed) significance of $1.9\sigma$ ($2.8\sigma$)\footnote{A cross-check using the dijet mass distribution yielded an expected (observed) significance of $1.4\sigma$ ($2.2\sigma$).}. All the other channels have undergone the same reinterpretation, with the goal to combine the results of these six channels.

\vspace*{-.1in}
\section{Results and future prospects}
\vspace*{-.1in}

\begin{figure}
\centering
\subfigure[$VH\to\met$\bbbar]{\includegraphics[width=4.5cm]{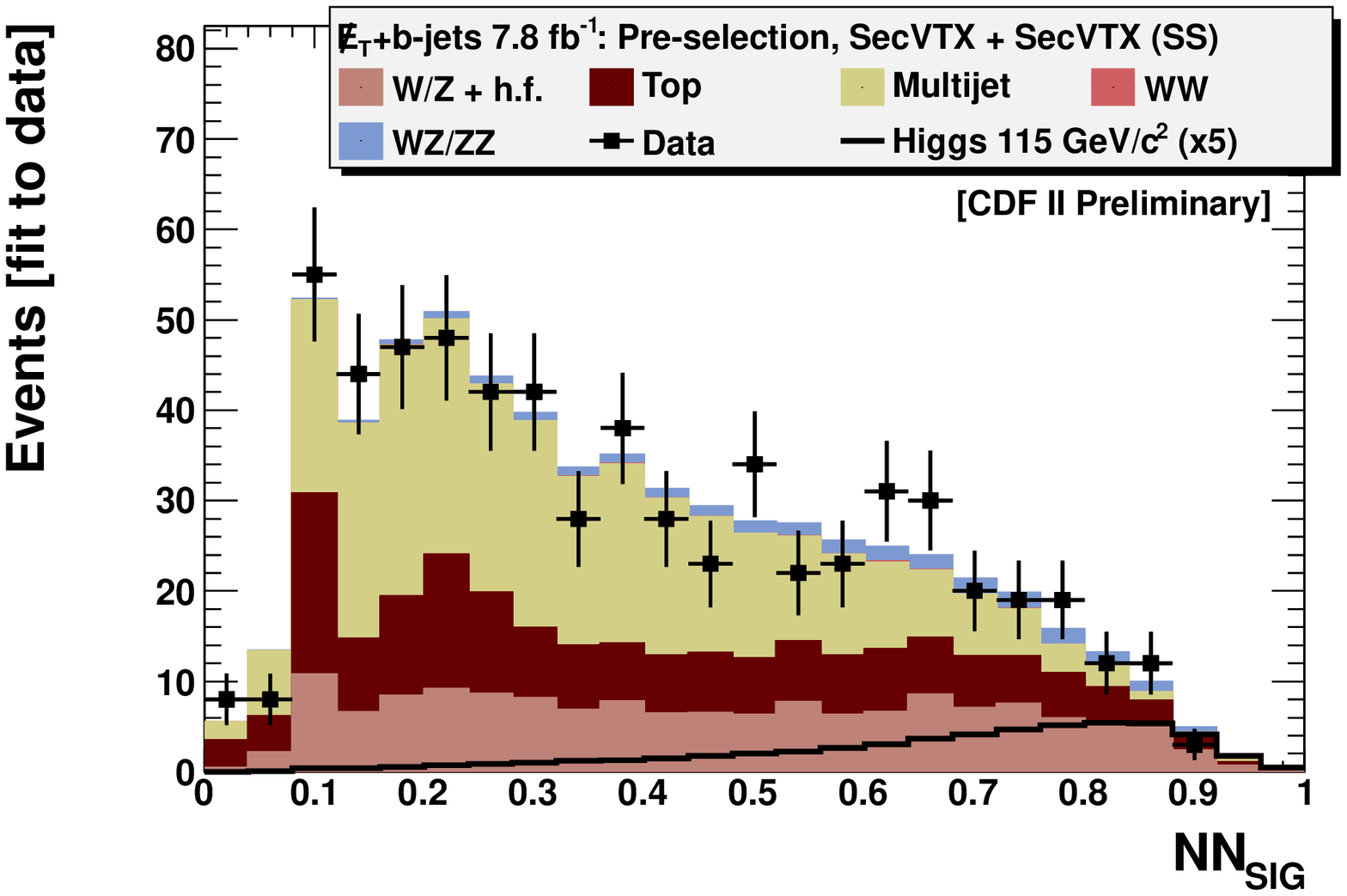}}
\subfigure[$WH\to\ell\nu$\bbbar]{\includegraphics[width=4.5cm,height=3.2cm]{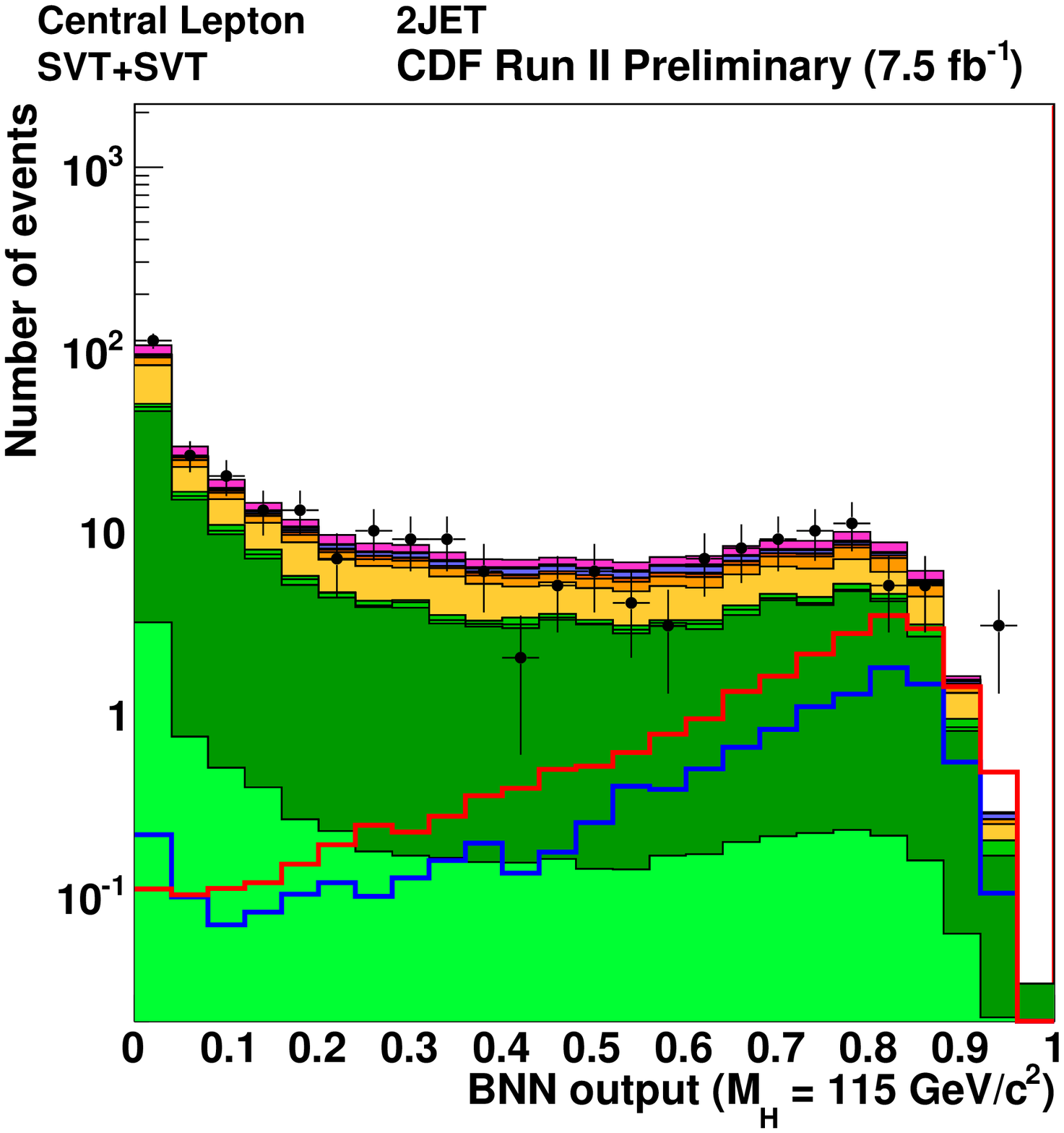}}
\subfigure[$ZH\to\ell\ell$\bbbar]{\includegraphics[width=4.5cm]{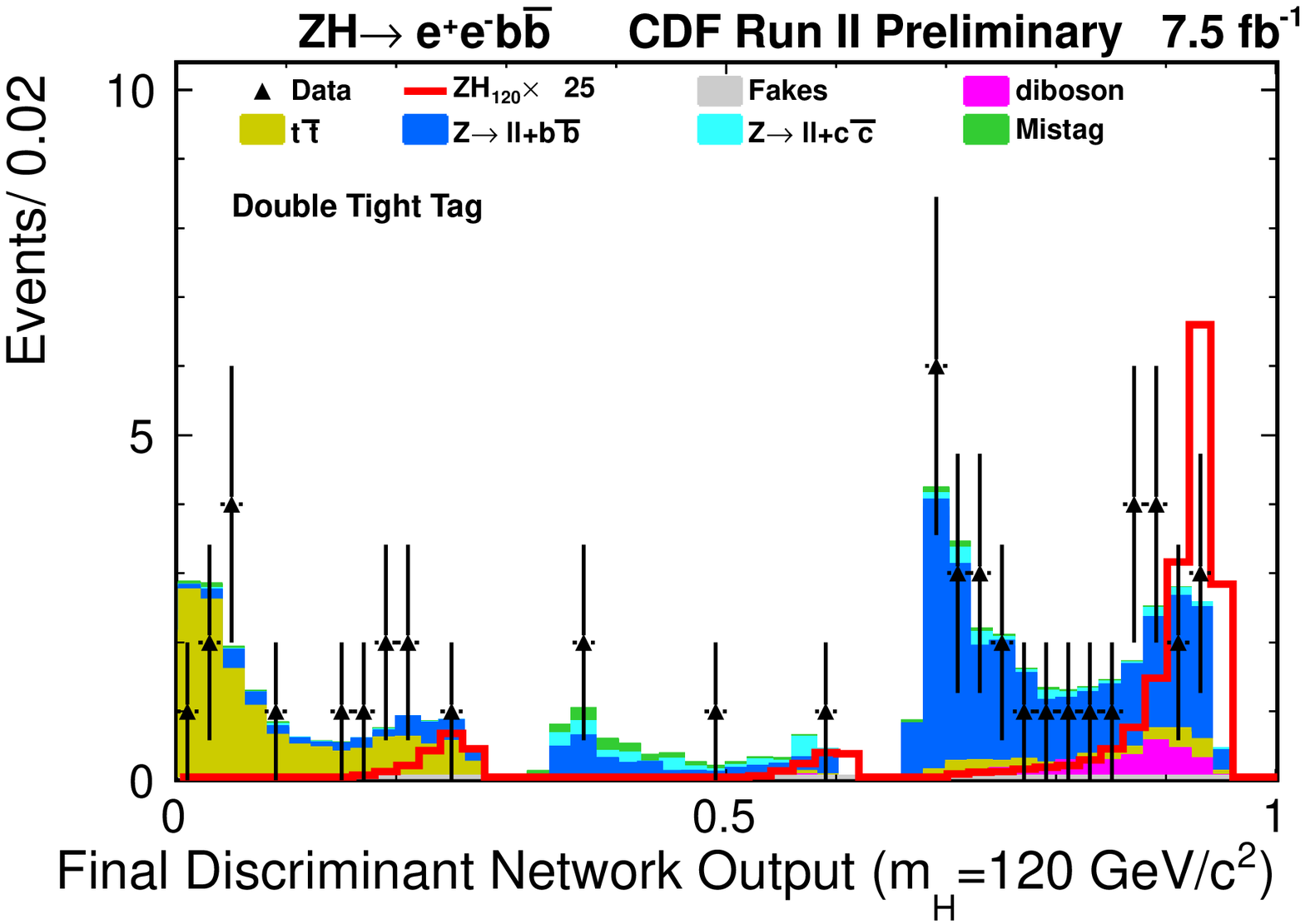}}
\caption{Distribution of the discriminant output for double tag events for each CDF analysis.}
\label{fig:cdf_LowMassHiggsResults}
\end{figure}

\begin{figure}
\centering 
\subfigure[$ZH\to\met$\bbbar]{\includegraphics[width=.3\linewidth,height=.2\linewidth]{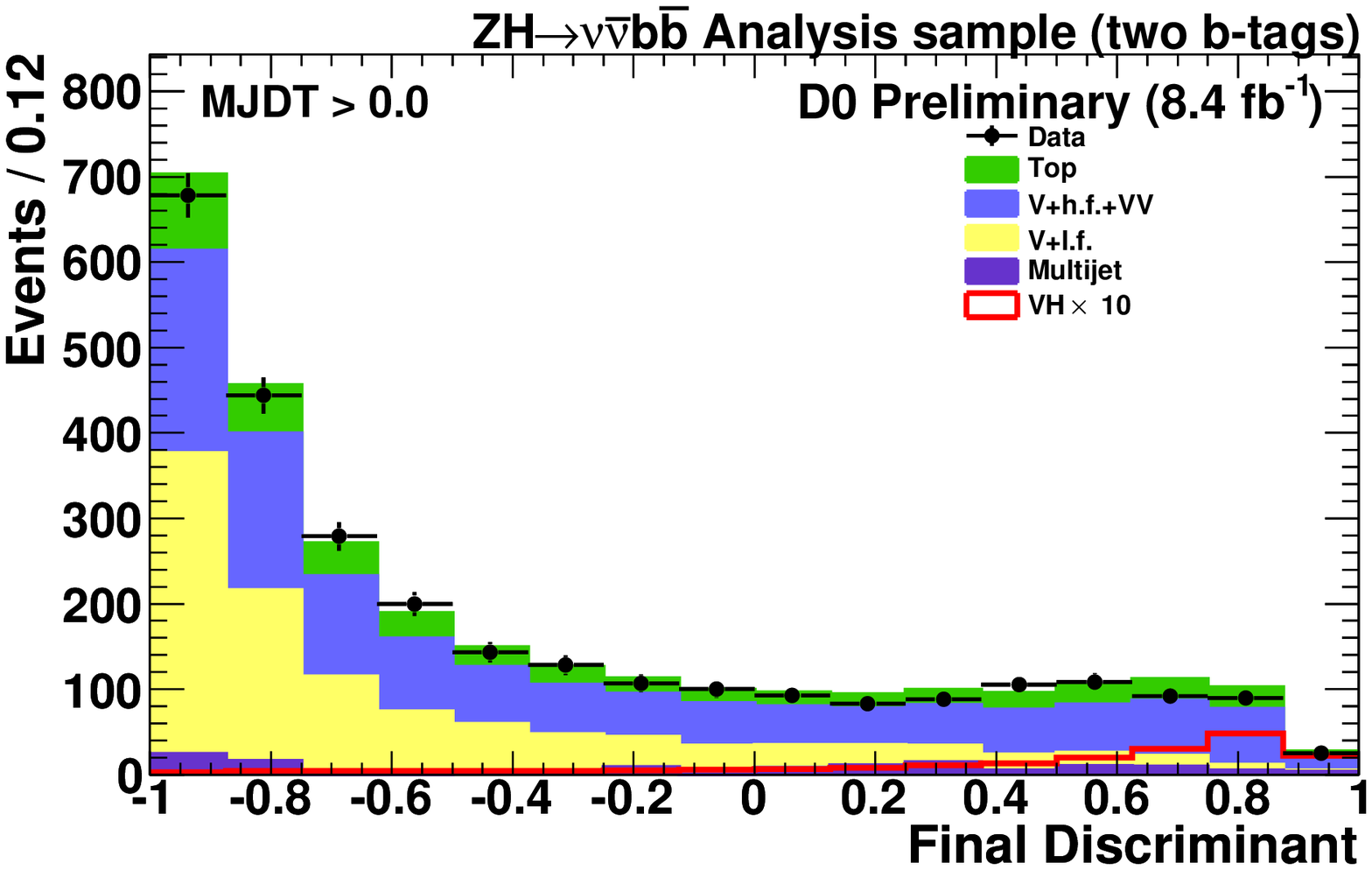}}
\subfigure[$WH\to\ell\nu$\bbbar]{\includegraphics[width=.3\linewidth,height=.2\linewidth]{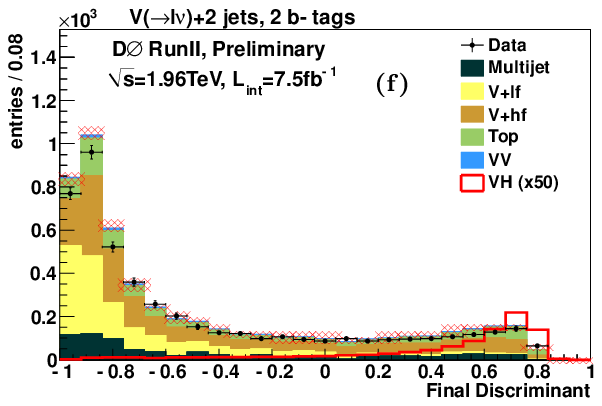}}
\subfigure[$ZH\to\ell\ell$\bbbar]{\includegraphics[width=.3\linewidth,height=.2\linewidth]{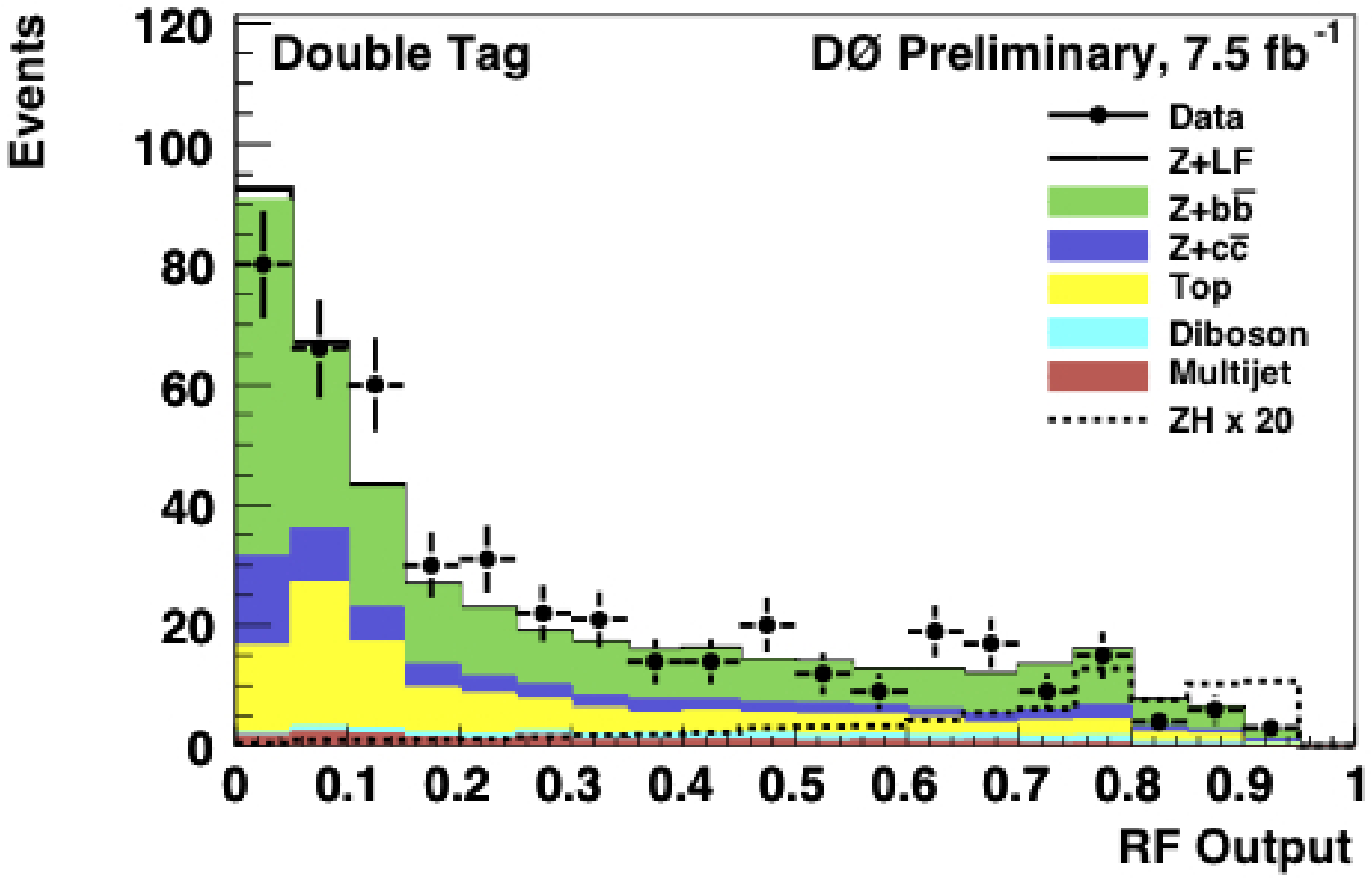}}
\caption{Distribution of the discriminant output for double tag events for each D\O\ analysis.}
\label{fig:dzero_LowMassHiggsResults}
\end{figure}

Figures~\ref{fig:cdf_LowMassHiggsResults} and ~\ref{fig:dzero_LowMassHiggsResults} show the distribution of the final discriminants for double tag events for the six analysis channels. 
Tables~\ref{tab:cdf_results} and~\ref{tab:dzero_results} show the expected and observed limits on the $VH$ production cross section relative to the SM expectation set by each analysis as a function of $m_H$.
For a mass of $115$~\gevcc, the best individual channel sets a 95\% C.L. limit of 2.3 times the SM expectation, with all limits below 5$\times$SM. The D\O\ collaboration updated the signal cross sections and branching ratios more recent predictions~\cite{ref:dzero_higgsXSec}, resulting in a loss of $\sim7\%$ in the predicted signal yields~\cite{ref:dzero_vvbb}.

We have presented the latest iterations of the three main low mass searches for a SM Higgs boson at CDF and D\O. Improved analysis techniques have allowed for yet another increase in sensitivity.  A $WZ+ZZ$ cross-section measurement is performed in each channel to validate the analysis technique of the Higgs search. Both collaborations will update these results with the final Tevatron dataset.


\begin{table}
\caption{\label{tab:cdf_results}The observed and expected upper limits measured using up to $7.8$~\invfb of CDF data on the $VH$ production cross section relative to the SM expectation as a function of $m_H$.}
\centering
\small
\begin{tabular}{ll*{11}{c}}
\multicolumn{2}{l}{$m_H$} & 100 & 105 & 110 & 115 & 120 & 125 & 130 & 135 & 140 & 145 & 150 \\\hline
\multirow{2}{*}{$ZH\to\met$\bbbar} & Exp. & 2.3 & 2.4 & 2.6 & 2.9 & 3.4 & 4.0 & 4.9 & 6.5 & 8.7 & 13.3 & 20.9 \\
& Obs. & 1.8 & 1.8 & 2.2 & 2.3 & 3.3 & 5.4 & 5.0 & 8.0 & 11.6 & 16.7 & 30.4  \\
\hline
\multirow{2}{*}{$WH\to\ell\nu$\bbbar} & Exp. & 1.8 & 2.0 & 2.2 & 2.6 & 3.1 & 3.7 & 4.8 & 6.4 & 8.8 & 14.2 & 21.6 \\
& Obs. & 1.1 & 2.1 & 2.8 & 2.7 & 3.4 & 4.4 & 6.1 & 7.7 & 12.3 & 18.9 & 34.4 \\
\hline
\multirow{2}{*}{$ZH\to\ell\ell$\bbbar} & Exp. & 2.7 & 3.1 & 3.4 & 3.9 & 4.7 & 5.5 & 7.0 & 8.7 & 11.9 & 17.5 & 27.7 \\
& Obs. & 2.8 & 3.3 & 4.4 & 4.8 & 5.4 & 4.9 & 6.6 & 7.4 & 10.3 & 13.8 & 21.8 \\
\end{tabular}
\end{table}

\begin{table}
\caption{\label{tab:dzero_results}The observed and expected upper limits measured using up to $8.5$~\invfb of D\O\ data on the $VH$ production cross section relative to the SM expectation as a function of $m_H$.}
\centering
\small
\begin{tabular}{ll*{11}{c}}
\multicolumn{2}{l}{$m_H$} & 100 & 105 & 110 & 115 & 120 & 125 & 130 & 135 & 140 & 145 & 150 \\\hline
\multirow{2}{*}{$ZH\to\met$\bbbar} & Exp. & 2.8 & 2.9 & 3.1 & 4.0 & 4.5 & 5.4 & 6.9 & 9.4 & 13.1 & 19.6 & 30.5 \\
& Obs. & 2.6 & 2.4 & 2.4 & 3.2 & 3.9 & 5.0 & 7.5 & 7.1 & 11.7 & 18.9 & 30.6 \\
\hline
\multirow{2}{*}{$WH\to\ell\nu$\bbbar} & Exp. & 2.4 & 2.6 & 3.0 & 3.5 & 4.3 & 5.4 & 7.0 & 9.6 & 13.6 & 20.4 & 33.6 \\
& Obs. & 2.6 & 2.9 & 4.1 & 4.6 & 5.8 & 6.8 & 8.2 & 6.3 & 10.3 & 13.3 & 23.2 \\
\hline
\multirow{2}{*}{$ZH\to\ell\ell$\bbbar} & Exp. & 3.4 & 3.7 & 4.2 & 4.8 & 5.3 & 6.5 & 8.4 & 11 & 14 & 22 & 34 \\
& Obs. & 2.5 & 2.6 & 3.1 & 4.9 & 6.4 & 8.9 & 9.9 & 15 & 25 & 34 & 50 \\
\end{tabular}
\end{table}


\clearpage


\begin{thebibliography}{99}
\bibitem{ref:BEHHGK}
Englert,~F. and Brout,~R. (1964),  {\em Phys. Rev. Lett.\/} {\bf 13} 321--323 ;
Higgs,~P.W.  (1964),  {\em Phys. Rev. Lett.\/} {\bf 13} 508--509 ;
Guralnik,~G.S. and Hagen,~C.R. and Kibble,~T.W.B,  (1964).  {\em Phys. Rev. Lett.\/} {\bf 13} 585--587.

\bibitem{ref:LEPEWWG}
Barate,~R. {\it et al.}, LEP Higgs Working Group (2003), {\em Phys. Lett. B\/} {\bf 565} 61--75 ;
\url{http://lepewwg.web.cern.ch/LEPEWWG/}

\bibitem{ref:single_top}
Abazov,~V.M. {\it et al.}, D\O\ Collaboration (2009), {\em Phys. Rev. Lett.\/} {\bf 103} 092001 ;
Aaltonen~T. {\it et al.}, CDF Collaboration (2009), {\em Phys. Rev. Lett.\/} {\bf 103} 092002.

\bibitem{ref:diboson}
Abazov,~V.M. {\it et al.}, D\O\ Collaboration (2009), {\em Phys. Rev. Lett.\/} {\bf 102} 161801 ;
Aaltonen~T. {\it et al.}, CDF Collaboration (2009), {\em Phys. Rev. Lett.\/} {\bf 103} 091803.


\bibitem{ref:CDFD0}
Acosta,~D.E. {\it et al.}, CDF Collaboration (2005), {\em Phys. Rev. D\/} {\bf 71} 032001; 
Abazov~V.M {\it et. al.}, Nucl. Instrum. Meth. A 565, 463 (2006).

\bibitem{ref:cdf_vvbb}
The CDF Collaboration, CDF note 10583,\\ \url{http://www-cdf.fnal.gov/physics/new/hdg/results/vhmetbb_110708/}

\bibitem{ref:dzero_vvbb}
The D\O\ Collaboration, D0 Note 6223-CONF,\\ \url{http://www-d0.fnal.gov/Run2Physics/WWW/results/prelim/HIGGS/H110/}

\bibitem{ref:cdf_lvbb}
The CDF Collaboration, CDF note 10596,\\ \url{http://www-cdf.fnal.gov/physics/new/hdg/results/whlnubb_071511/}

\bibitem{ref:dzero_lvbb}
The D\O\ Collaboration, D0 Note 6220-CONF,\\ \url{http://www-d0.fnal.gov/Run2Physics/WWW/results/prelim/HIGGS/H108/}

\bibitem{ref:cdf_llbb}
The CDF Collaboration, CDF notes 10593 and 10572,\\ \url{http://www-cdf.fnal.gov/physics/new/hdg/results/zhllbb_comb_110715/}


\bibitem{ref:dzero_llbb}
The D\O\ Collaboration, D0 Note 6166-CONF,\\ \url{http://www-d0.fnal.gov/Run2Physics/WWW/results/prelim/HIGGS/H109/}

\bibitem{ref:tevHiggsComb2011}
The CDF \& D\O\ Collaborations, the Tevatron New Phenomena Higgs Working Group,\\
CDF note 10606, D\O\ Note 6226, FERMILAB-CONF-11-354-E, arXiv:1107.5518 [hep-ex],\\ \url{http://tevnphwg.fnal.gov/results/SM_Higgs_Summer_11/}

\bibitem{ref:cdf_secvtx}
Acosta,~D.E. {\it et al.}, CDF Collaboration (2005), {\em Phys. Rev. D\/} {\bf 71} 052003

\bibitem{ref:cdf_jetprob}
Abulencia,~A. {\it et al.}, CDF Collaboration (2006), {\em Phys. Rev. D\/} {\bf 74} 072006 

\bibitem{ref:dzero_btagging}
Abazov~V.M {\it et. al.}, Nucl. Instrum. Meth. A 620, 400 (2010).

\bibitem{ref:dzero_higgsXSec}
J. Baglio and A. Djouadi, arXiv:1003.4266v2 [hep-ph].

\end{thebibliography}
\end{document}